\documentclass[conference]{IEEEtran}
\IEEEoverridecommandlockouts
\usepackage{cite}
\usepackage{amsmath,amssymb,amsfonts}
\usepackage{algorithmic}
\usepackage{graphicx}
\usepackage{textcomp}
\usepackage{booktabs}
\usepackage{xcolor}
\def\BibTeX{{\rm B\kern-.05em{\sc i\kern-.025em b}\kern-.08em
    T\kern-.1667em\lower.7ex\hbox{E}\kern-.125emX}}
\begin{document}

\title{A comparative study of sensory encoding models for human navigation in virtual reality}

\author{Tangyao Li, Qiyuan Zhan, Yitong Zhu, Bojing Hou, and Yuyang Wang
\thanks{Corresponding author. Email: yuyangwang@hkust-gz.edu.cn}}

\maketitle

\begin{abstract}

In virtual reality applications, users often navigate through virtual environments, but the issue of physiological responses, such as cybersickness, fatigue, and cognitive workload, can disrupt or even halt these activities. Despite its impact, the underlying mechanisms of how the sensory system encodes information in VR remain unclear. In this study, we compare three sensory encoding models, Bayesian Efficient Coding, Fitness Maximizing Coding, and the Linear Nonlinear Poisson model, regarding their ability to simulate human navigation behavior in VR. By incorporating the factor of physiological responses into the models, we find that the Bayesian Efficient Coding model generally outperforms the others. Furthermore, the Fitness Maximizing Code framework provides more accurate estimates when the error penalty is small. Our results suggest that the Bayesian Efficient Coding framework offers superior predictions in most scenarios, providing a better understanding of human navigation behavior in VR environments.

\end{abstract}


\section{Introduction}
Virtual Reality (VR) is a rapidly expanding field with numerous applications across various domains, thanks to its ability to create controlled, reproducible environments \cite{yaremych2019tracing}. Virtual environments provide users with unique experiences, but the sensory conflict between visual stimuli and physical sensations can cause discomfort, often presenting as symptoms of cybersickness \cite{reason1975motion}. Beyond discomfort, navigating VR can increase cognitive workload and fatigue, which in turn affects user task performance \cite{sheik2007utility,wang2021enhanced}. Therefore, it is essential to comprehend how to simulate and anticipate physiological responses to these elements to overall user experiences.

Human physiological signals could provide knowledge about one’s internal state, such as level of discomfort, cognitive workload, and fatigue \cite{critchley2002electrodermal}. During immersion in the virtual environment, these signals could serve as an indicator of one’s engagement and body condition \cite{carlson2012physiology}. Nonetheless, understanding how these signals emerge during virtual navigation poses a significant challenge \cite{arnone2011curiosity}. This study seeks to investigate and assess several computational models that aim to replicate the dynamics of these physiological reactions during virtual navigation.

We focus on three computational models: Bayesian Efficient Coding (BEC), Fitness Maximizing Code (FMC), and Linear-Nonlinear Poisson (LNP). We selected these models because they can effectively capture the various aspects of the user's navigation strategy. The BEC model illustrates how the brain minimizes uncertainty, relevant as users navigate the virtual space with limited information. The FMC model highlights the trade-offs between navigation and managing discomfort, including fatigue, cybersickness, and stress. Furthermore, a frequently utilized model for simulating the relationship between sensory input and neural response is the LNP model. Employing this model can effectively represent how users react to stimulus encounters during immersion.

We simulate the relationship between user navigation data (i.e., motion/speed profile) and their corresponding physiological responses. By simulating user response and comparing the results to the principle of maximum information preservation, which is widely regarded as an optimal encoding strategy for sensory information, we assess the effectiveness of each model in capturing the key factors influencing user interaction in VR. This paper presents the following contributions:

\begin{itemize}
    \item We analyze how different models can be used to predict the underlying physiological responses during virtual navigation.
    \item We explore how these models can generalize to various aspects of user experience in virtual environments, shedding light on the broader potential for improving VR interaction and usability.
\end{itemize}

\section{Related Works}

\subsection{Human Behavior in VR}

Behavioral differences in movement patterns are evident in virtual environments. As shown in \cite{deb2017efficacy}, walking speeds in VR are generally similar to those in the real world. However, Iryo-Asano, Hasegawa and Dias \cite{iryo2018applicability} demonstrated that the average maximum walking speed is lower in VR, likely due to users maintaining larger safety distances, which stem from the limited field of view provided by head-mounted displays (HMDs). Additionally, users struggle with spatial awareness in VR, as they cannot see their own bodies, making it harder to judge proximity to virtual objects.

These differences in user behavior are influenced by the unique sensory inputs and immersive nature of VR. For example, restricted visual feedback and limited proprioception often result in slower movement in VR \cite{deb2017efficacy}, and the type of input device, such as HMDs or motion tracking, can affect spatial awareness and movement accuracy \cite{riecke2010we}. Understanding these factors is essential for designing effective VR interfaces.

Moreover, prior experience with virtual environments plays a significant role in user behavior. Smith and Du'Mont \cite{Smith2009} showed that gamers, accustomed to rapid and precise movements, perform better in VR tasks, including spatial perception, task completion time, and movement accuracy. These skills are transferable, enabling experienced users to adapt quickly and maximize comfort in VR environments.

\subsection{Physiological Responses in VR}

During virtual navigation, users may experience a range of physiological responses that can influence their ability to navigate and interact. Physiological signals, such as heart rate, galvanic skin response, and eye movement, are commonly used to measure different physiological responses (cybersickness \cite{venkatakrishnan2020comparative,li2023deep}, motion \cite{egger2019emotion}, stress \cite{zhai2008stress}, and cognitive workload \cite{martens2019feels}) during virtual immersion. The Galvanic Skin Response (GSR) signal measures the skin's electrical conductance, which varies with sweat gland activity under the control of the sympathetic nervous system \cite{alexander2005separating}. GSR can be used to indicate emotion, sickness, and stress response, allowing us to understand how the stimulus affects the participant's condition. From the GSR signal (also known as Electrodermal Activity (EDA)), the tonic and phasic components can be extracted, which represent long-term physiological states such as general arousal or stress, and immediate physiological reactions to stimuli, respectively.

Cybersickness has been shown to significantly impair both cognitive and motor performance in immersive VR environments \cite{10.3389/fnhum.2020.00096}. It arises when there is a mismatch between visual sensory input and proprioceptive feedback, often leading to symptoms like nausea and dizziness \cite{reason1975motion}.  However, the findings on this issue are not entirely consistent. Supporting the view of temporary cognitive decline. As shown in \cite{Dahlman-Joakim}, motion sickness notably reduces users' verbal working memory.

Moreover, the user is required to process a substantial amount of sensory information during virtual immersion. This continuous stream of sensory data may increase an individual's cognitive load \cite{tremmel2019estimating}. Research has shown that high cognitive load can make it more difficult for users to employ explicit strategies for adapting to and learning new tasks effectively \cite{martens2019feels}. Consequently, in order to accurately model sensory processing, it is important to comprehend the impact of cognitive workload. During virtual navigation, several factors could heighten an individual's cognitive load. For example, time constraints and task complexity increase the user's cognitive load, which in turn can trigger a stress response. Furthermore, as cognitive workload increases, mental stress is likely to escalate and impaire working memory as well. In essence, when users are faced with overwhelming cognitive challenges, their capacity to process information, make decisions, and perform tasks effectively diminishes.

\subsection{Sensory Encoding Theories}

The Efficient Coding Hypothesis (ECH) provides a theoretical framework for understanding sensory encoding, proposing that sensory systems encode information in a manner that maximizes efficiency, typically by minimizing redundancy \cite{barlow1961possible}. In this context, redundancy refers to the portion of the sensory channel that is not used to transmit useful information. The Infomax (principle of maximum information preservation) coding \cite{linsker1988self}, a specific strategy within the ECH, aims to transmit as much information as possible by maximizing the mutual information between the sensory input and the neural representation \cite{park2017bayesian}. However, there is an ongoing debate on whether the primary goal of sensory systems is to achieve high accuracy while maintaining efficiency, or whether other factors, such as adaptability or robustness, play a more significant role in sensory processing \cite{schaffner2023sensory}.

While the ECH emphasizes the efficiency, Bayesian inference provides a probabilistic framework for sensory processing. This approach employs Bayes’ theorem to revise the probability of a hypothesis based on new observational data \cite{dempster1968generalization}. Bayes' theorem states that the posterior probability of a hypothesis is obtained by merging the likelihood of observed data with the prior probability, which represents the initial belief before making the observation..

In contrast to both the ECH and Bayesian inference, the FMC proposes a different theoretical perspective. This approach focuses on how sensory processing and decision-making can maximize accuracy (or equivalently, minimize error), while also considering the long-term outcomes, such as maximizing overall reward or fitness over time \cite{schaffner2023sensory}. By emphasizing the adaptive response of organisms to their environment, the FMC highlights the importance of fitness in shaping sensory processing strategies.

\section{Methods}
\subsection{Bayesian Efficient Coding}
Bayesian Efficient Coding (BEC) is a theoretical framework that combines the ECH \cite{barlow1961possible} with Bayesian inference \cite{dempster1968generalization} to explain how sensory systems optimize the encoding of stimuli while adhering to constraints on neural resources \cite{park2017bayesian}. The objective is to minimize the uncertainty in sensory encoding by reducing posterior entropy. The posterior entropy represents the uncertainty that remains after observing the neural response to a particular stimulus. While classical efficient coding theory focuses primarily on maximizing mutual information, BEC extends this framework by incorporating a broader range of loss functions. This added flexibility allows for the exploration of different neural coding strategies beyond the traditional goal of mutual information maximization \cite{oliver1952efficient}. In this work, we define the loss function for the BEC model as
\begin{equation}
    Loss_{BEC} (\theta,p) = \sum_{i=1}^{n}\big( y_{noise,i} - g_{1}(x_{std,i},\theta)\big)^{p},
\end{equation}
where
\begin{itemize}
    \item $g_{1}(x_{std,i};\theta)=\frac{1}{2} \cdot \big(1+\tanh(\theta_{1}\cdot x_{std,i}+\theta_{2})\big)$ is the nonlinearity that represents LMC response with random initialized parameter $\theta=[\theta_{1},\theta_{2}]$ and stimulus $x_{std,i}$,
    \item $y_{noise,i}=g(x_{std};\theta)+\epsilon$ is the LMC response with Gaussian noise $\epsilon \sim \mathcal{N}(0,0.05^{2})$,
    \item $p$ is the error penalty.
\end{itemize}
The goal for the BEC method is to find the optimal neural encoding strategy that minimizes the loss function $Loss_{BEC}$ subject to the capacity constraint on neural response \cite{park2017bayesian}.

\subsection{Fitness Maximizing Code}
Fitness Maximizing Code (FMC) focuses on how organisms can maximize their fitness or reproductive success through coding strategies. It may prioritize maximizing the total reward or fitness gained by the organism over the long term instead of maximizing information transfer (e.g., Infomax coding) or minimizing errors (e.g., BEC) \cite{schaffner2023sensory}. In this work context, the concept of fitness is associated with the user’s physiological position, including fatigue, cybersickness, stress, and cognitive workload.

The loss function for the FMC method is
\begin{equation}
    Loss_{FMC} (\theta,p) = \sum^{n}_{i=1} \Big( w(y_{noise,i}) \cdot \big(y_{noise,i}-g_{1}(x_{std,i},\theta) \big)^{p} \Big),
\end{equation}
where $w(y_{noise,i})=|y_{noise,i}|$ reprsents the weight.

\subsection{Linear Nonlinear Poisson Model}
The Linear Nonlinear Poisson (LNP) model \cite{simoncelli2004characterization} is a widely used framework in computational neuroscience for modeling neural responses to stimuli. Unlike the BEC method, which focuses on optimizing the encoding of sensory information, the LNP model focuses on modeling the dynamics of sensory neuron responses. It is particularly effective for describing neural responses characterized by stochastic spiking behavior, a common feature in sensory systems \cite{corrado2005linear}. The LNP model consists of three stages:
\begin{enumerate}
    \item Linear stage: The stimulus $x_{std,i}$ is linearly filtered through a receptive field or a set of filters, which can be represented as a weighted sum of the stimulus components.
    \item Nonlinear stage: The output of the linear filter is then passed through a nonlinear function $g_{2}(x_{std,i})$, which accounts for the rectification, saturation, or other nonlinear transformations observed in neural responses.
    \item Poisson stage: The output of the nonlinear stage is used as the rate parameter of a Poisson process, governing the probability of generating spikes in response to the stimulus.
\end{enumerate}
We minimized the loss function for the LNP model
\begin{equation}
    \begin{aligned}
        Loss_{LNP} (\theta,p) &= \sum^{n}_{i=1} |y_{i}-g_{2}(x_{std,i},\theta)|^{p} \\
        &+ w_{penalty} \Big( |\overline{y_i}-\frac{1}{n}\sum^{n}_{i=1}g_{2}(x_{std,i},\theta)| \\
        &+ \Delta^{2}g_{2}(x_{std,i},\theta) \Big),
    \end{aligned}
\end{equation}
where
\begin{itemize}
    \item $\overline{y_i}$ is the mean value,
    \item $w_{penalty}$ represents the penalty weight,
    \item $g_{2}(x_{std,i},\theta)=\max(0,\theta_{0}x_{std,i}^{2}+\theta_{1}x_{std,i}+\theta_{2})$ for the random initialized parameter $\theta=[\theta_{0}, \theta_{1},\theta_{2}]$,
    \item $\Delta^{2}g_{2}(x_{std,i},\theta)$ is the second-order difference of the function $g_{2}(x_{std,i},\theta)$.
\end{itemize}

\subsection{Data Processing}
\begin{figure}[htbp]
    \centering
    \includegraphics[width=0.9\linewidth]{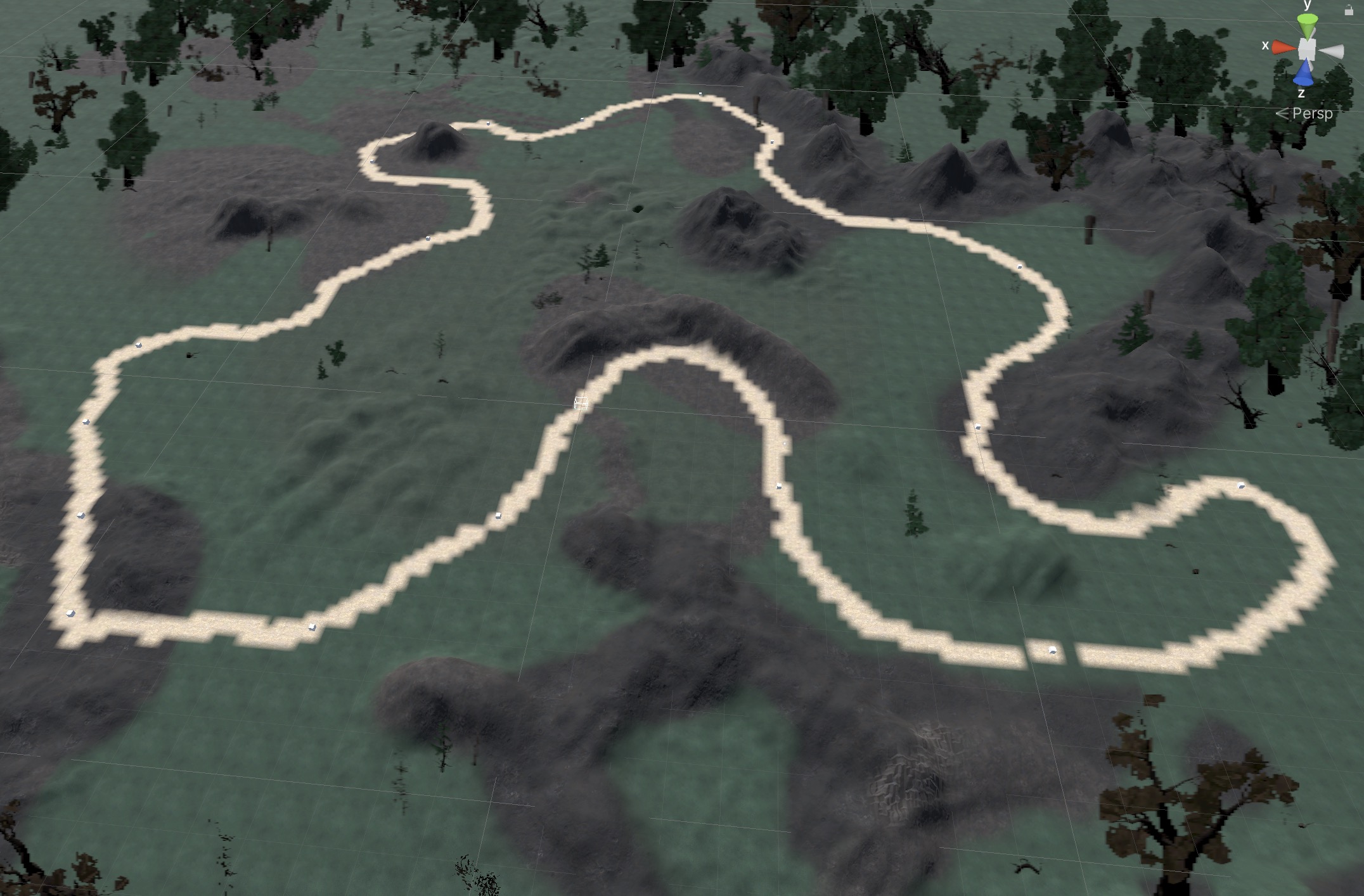}
    \caption{The virtual environment in which the participant carries out the navigation task. The participant navigated along a highlighted path in the virtual environment, which guided their movements throughout the task.}
    \label{fig:navigation}
\end{figure}

We utilized a dataset \cite{wang2023dataset} that collects user motion and biosignal data during a virtual navigation task (see Table \ref{tab:data} and Fig. \ref{fig:navigation}). The participants were equipped with an HTC Vive Pro headset and an Empatica E4 wristband,  which recorded physiological signals, including GSR, blood volume pulse, temperature, and heart rate. This dataset includes 53 participants' navigation data, with 26 females and 27 males. The average age among the participants is 26.3 years, with a standard deviation of 3.3.

The participant's navigation speed $x$ is standardized using
\[x_{std} = \frac{x - \mu_{x}}{\sigma_{x}}.\]
We extracted the tonic and phasic components from the user's GSR data, where the former represents general physiological states that appeared without external stimulus, and the latter reflects the immediate physiological reactions to stimuli. We then use the phasic EDA signal to represent the ground truth value of the participant's physiological state during virtual immersion.

\begin{table}[htbp]
    \caption{Summary of data}
    \begin{center}
    \begin{tabular}{c|l}
    \toprule
    \textbf{Type} & \textbf{Data} \\
    \hline
    & Raw speed \\
    Motion data & Resampled data \\
    & Raw rotation \\
    & Resampled rotation \\
    \hline
    & Galvanic skin reponse (4Hz) \\
    Biosignal data & Blood volume pulse (64Hz) \\
    & Temperature (4Hz) \\
    & Heart rate \\
    \bottomrule
    \end{tabular}
    \label{tab:data}
    \end{center}
\end{table}

\section{Results}
In this section, we present the results for the optimized responses with three methods from one participant's data. The Infomax response has been included to provide a comparable benchmark for reference across the three methods.

\subsection{BEC Model}

\begin{figure}[htbp]
    \centering
    \includegraphics[width=\linewidth]{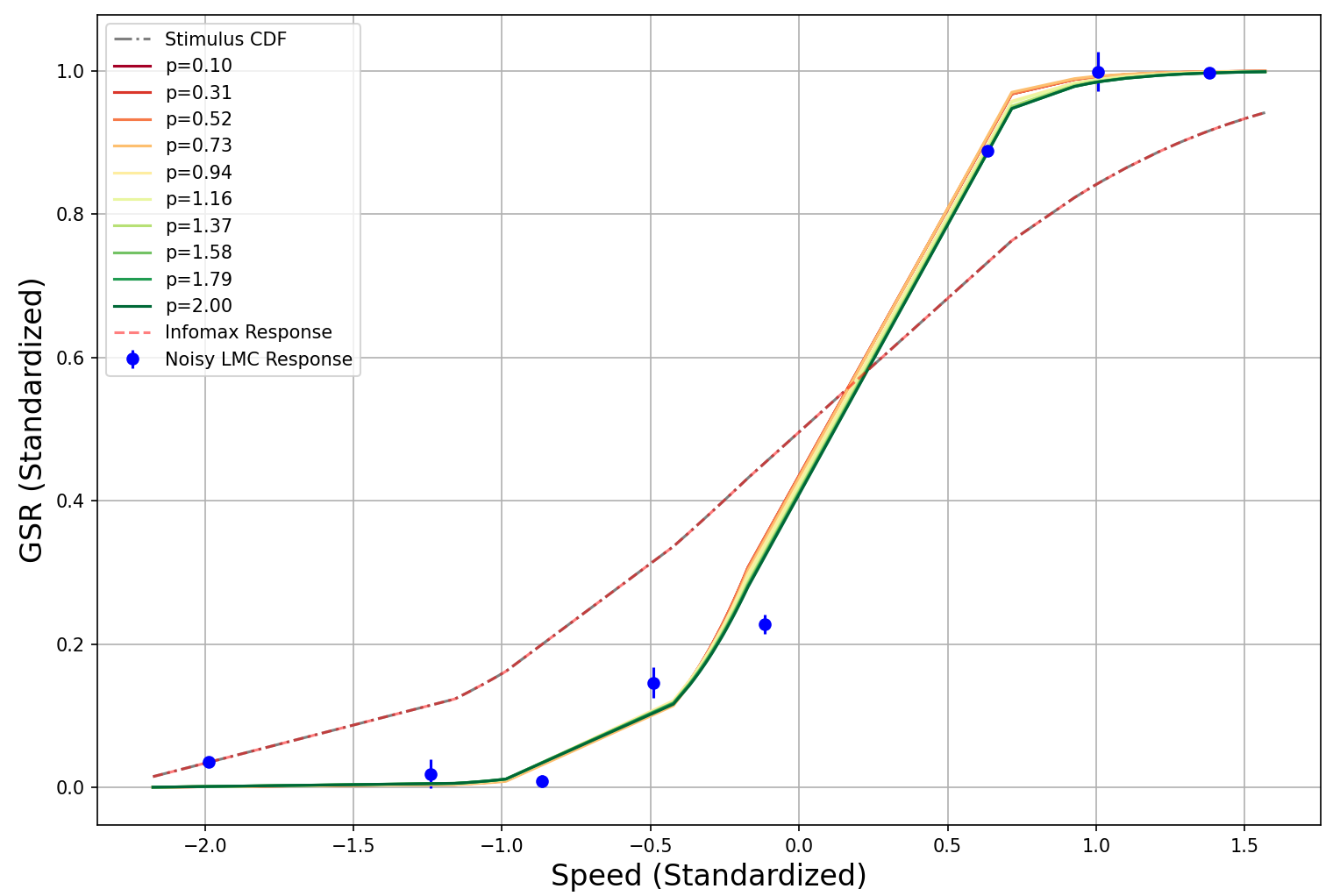}
    \caption{Cumulative distribution function of the stimulus and the optimized large monopolar cell response (with $p$ = 0.10 to 2.00) for a single participant using the BEC model. The red dashed curve represents the Infomax response derived from the stimulus cumulative distribution function (shown in grey). The blue dots represent the LMC response.}
    \label{fig:BEC}
\end{figure}

The stimulus cumulative distribution function (CDF) with different error penalty $p$ is shown in Fig. \ref{fig:BEC}. From the figure, we can observe that although there is a small noticeable difference around the standardized speed 0 -- 0.05 m/s, there is no significant difference between the CDF curves for different values of $p$. Moreover, the trend of the optimized responses closely resembles that of the Infomax response and the stimulus CDF, with the Infomax response curve being almost identical to the stimulus CDF curve.

\subsection{BEC with FMC loss}
\begin{figure}[htbp]
    \centering
    \includegraphics[width=\linewidth]{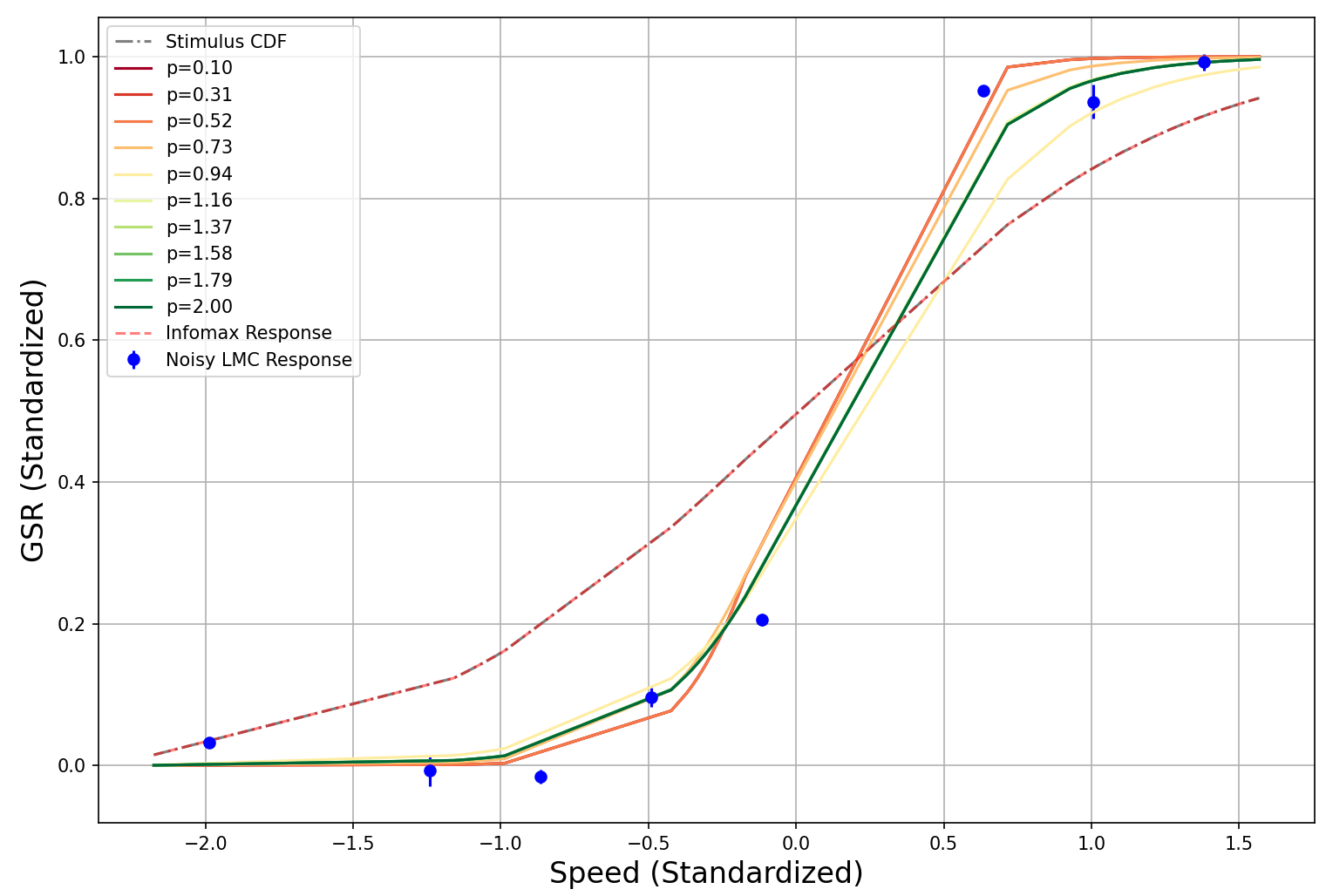}
    \caption{Cumulative distribution function of the stimulus and the optimized large monopolar cell response (with $p$ = 0.10 to 2.00) for a single participant using the BEC model with the FMC loss function. The red dashed curve represents the Infomax response derived from the stimulus cumulative distribution function (shown in grey). The blue dots represent the LMC response.}
    \label{fig:FMC}
\end{figure}

Fig. \ref{fig:FMC} shows the results for a single participant using the BEC with the FMC loss function model at various $p$ values. As in the BEC model, the Infomax response and the stimulus CDF curves are nearly identical in this plot. Similar to the result of the BEC model, we observe that there are no significantly differences in the trends of the optimized LMC responses across the different $p$ values, the responses do vary with $p$. Despite these variations, the optimized LMC responses follow a similar pattern to the Infomax response curve.

In contrast with the BEC result, we recognize a noticeable difference between all stimulus CDF curves starting from a standardized speed of -1.0m/s. Moreover, it can be observed that the difference between the stimulus CDF curves becomes more distinguishable around the standardized speed of 0.5 -- 1.0m/s.

\subsection{LNP Model}
\begin{figure}[htbp]
    \centering
    \includegraphics[width=\linewidth]{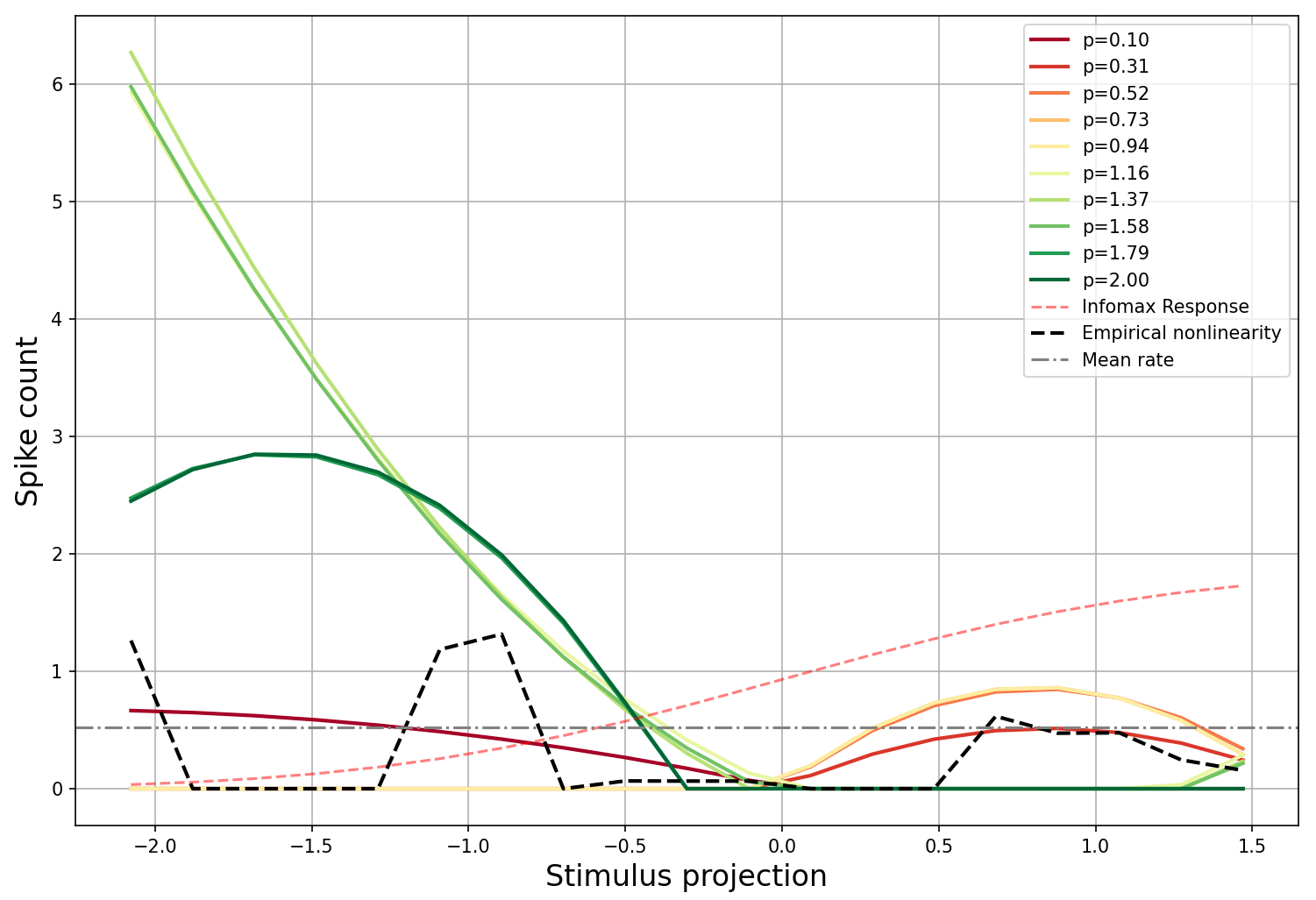}
    \caption{Empirical data and nonlinearity for one participant using the LNP model with different $p$ values. The black dashed curve indicates the Poisson-Gaussian process fit of the nonlinear rate function, while the red dashed curve represents the Infomax response.}
    \label{fig:LNP}
\end{figure}

The Infomax response, empirical nonlinearity, and optimal nonlinearities are shown in Fig. \ref{fig:LNP}. Overall, the optimal nonlinearities do not follow a comparable trend with the Infomax response. We observe that the optimal nonlinearities with $p=$ 0.31, 0.52, 0.73, and 0.94 follow a similar pattern to the empirical nonlinearity in the latter half. Among these, the optimal nonlinearity for $p=0.31$ most closely matches the empirical nonlinearity. However, in the first half of the empirical nonlinearity, it is difficult to identify an optimal nonlinearity that follows the same pattern. Of the optimal nonlinearities in the first half, the one for $p=$ 0.10 is the most similar to the empirical pattern. On the other hand, the MSE results (see Table \ref{tab:mse_result}) indicate that the optimal nonlinearity for $p=$ 2.00 is most similar to the Infomax response (MSE=0.055115).

\section{Discussion}

\begin{table}
    \caption{Mean square error result for the BEC, BEC with FMC loss, and LNP methods with different $p$ values.}
    \begin{center}
    \begin{tabular}{c|c|c|c}
        \toprule
        \textbf{$p$ value} & \textbf{BEC} & \textbf{BEC with FMC loss} & \textbf{LNP} \\ \hline
        0.10 & 0.002869 & 0.002801 & 0.369924 \\ \hline
        0.31 & 0.002869 & 0.002801 & 0.532142 \\ \hline
        0.52 & 0.002869 & 0.002801 & 0.527909 \\ \hline
        0.73 & 0.002779 & 0.002955 & 0.527733 \\ \hline
        0.94 & 0.002744 & 0.003158 & 0.527733 \\ \hline
        1.16 & 0.002666 & 0.002680 & 0.115103 \\ \hline
        1.37 & 0.002601 & 0.002690 & 0.122359 \\ \hline
        1.58 & 0.002576 & 0.002687 & 0.116662 \\ \hline
        1.79 & 0.002558 & 0.002683 & 0.055265 \\ \hline
        2.00 & 0.002541 & 0.002677 & 0.055115 \\ \bottomrule
    \end{tabular}
    \label{tab:mse_result}
    \end{center}
\end{table}

\subsection{Model Evaluation}

One of the main goals of this work is to evaluate the performance of each method. We introduced the Infomax response, which aims to maximize the mutual information between the input and output of a system \cite{park2017bayesian}, as a standard for comparison. To quantify the difference between the results, we compute the mean squared error (MSE) for each approach as a performance measure (see Table \ref{tab:mse_result}).

Firstly, we evaluate the suitability of the BEC model. The results suggest that the MSE remains consistently small for both the BEC model with only slight fluctuations across different $p$ values. Similarly, the performance of the BEC with the FMC loss model is stable according to the MSE results. At smaller error penalty values ($p < 0.73$), the combination of BEC and FMC outperforms the standard BEC model in terms of accuracy, as reflected in lower MSE values. However, for $p \geq 0.73$, the standard BEC model demonstrates better performance. A possible explanation for this might be that accuracy may be sacrificed to achieve better long-term fitness in the BEC with FMC model, aligning with the trade-offs proposed by the fitness-maximizing framework. Additionally, when $p$ is small, the error penalty on the difference between the predicted and actual results is relatively minimal. Therefore, the BEC with FMC loss model performs better than the BEC model. In contrast, as the penalty increases, the error grows more significant, thereby impacting accuracy and the performance switches at larger $p$ values. This finding is contrary to a previous study \cite{schaffner2023sensory}, which has reported that sensory processing encodes maximized fitness instead of accuracy at an early stage.

For one of the participants, her Simulator Sickness Questionnaire (SSQ) score was classified as ‘negligible’ (SSQ=3.74). This may imply that the user can explore the environment more freely since the SSQ value shows minimal discomfort, indicating a shift in the balance between exploration and discomfort management. In addition, the participant selects ‘often’ for her gaming experience, suggesting that previous gameplay may have aided her navigation task cooperation \cite{tian2022review,wang2021using}. We may imply that users with low discomfort or frequent gaming experience are more likely to engage in exploration and complete tasks without altering their behavior due to discomfort \cite{venkatakrishnan2020comparative}. Therefore, when discomfort is negligible, behavior aligns more with optimal exploration (i.e., reduce uncertainty). Conversely, higher discomfort levels result in a greater need for fitness-maximizing decisions, such as prioritizing decisions to minimize discomfort over maximizing exploration.

In VR environments, sensory conflicts between visual input and body perception are widely believed to contribute to cybersickness \cite{reason1975motion}. Our result suggests the BEC and FMC models can account for this sensory conflict, as their primary objectives are to reduce uncertainty (BEC) and optimize long-term fitness (FMC). Overall, both methods exhibit robustness across a range of $p$ values and achieve high levels of accuracy, although the BEC model generally outperforms the BEC with FMC loss under larger error penalties.

On the other hand, the LNP model shows slightly lower accuracy compared to the other methods. From Table \ref{tab:mse_result}, the MSE results suggest that its performance is not as good as that of the other two models. Moreover, the MSE values for the LNP model fluctuate more significantly across different $p$ values, suggesting that its performance is more sensitive to the choice of $p$. Furthermore, the fluctuating pattern suggests that the LNP model appears to perform better when the error penalty is larger (closer to 2). This indication highlights the importance of parameter selection in determining the model's accuracy.

Several factors may explain the relatively lower accuracy of the LNP model. Firstly, the LNP model starts by linearly filtering the input signal \cite{simoncelli2004characterization}. As a result, any nonlinear relationships between the signals may be lost in this process, potentially hindering the model's ability to capture the full complexity of the input data. Another possible explanation is the assumption that the user’s GSR signal is analogous to the spike counts used in the Poisson process may oversimplify the dynamics of neural activity. This simplification might prevent the LNP model from accurately capturing the complexities of the user’s sensory experience in VR, particularly in relation to discomfort or cybersickness.

\subsection{Limitations}
Our work has certain limitations. Initially, some fixed assumptions might not match the intricate and dynamic sensory inputs found in the VR environment, where user behavior is often unpredictable and constantly changing. A potential solution is to incorporate a mechanism that allows the model to adapt based on prior experiences and accumulated knowledge.

Another limitation lies in the simplification of the models. For example, the BEC model assumes that sensory information is encoded in a way that minimizes uncertainty while maximizing information transfer. However, this assumption may not fully capture the complexity of human perception during virtual navigation. Similarly, the FMC model assumes the brain strikes a balance between accuracy and fitness, which may not always hold true in real-world scenarios.

Despite these limitations, the models in this study provide important insights into how sensory systems might encode and process information in VR settings. Future research could tackle these limitations by integrating more complex sensory inputs, considering individual differences, and exploring real-time adaptation based on user feedback.

\section{Conclusion}
This study examines the suitability of modeling human navigation behavior in virtual reality using the BEC, FMC, and LNP models. We adapted the FMC loss function into the BEC model because the FMC method provides a relevant framework for virtual navigation. In this context, participants have make navigation decisions while considering their current physiological state. Specifically, when participants experience high levels of discomfort, they may reduce their movement. The LNP model was also considered because it models neural responses using a stochastic spiking process. After comparing the MSE value between the three approaches, we found that the BEC model provides a more accurate estimation for modeling human behavior during virtual navigation. While the model that combines BEC and FMC produces slightly less accurate predictions, the concept of maximizing fitness within this model reflects the trade-off between exploring the virtual environment and managing physiological conditions. In short, the results suggest that while each model has their advantages, the BEC model offers a more accurate framework for simulating human navigation behavior in VR.

\section*{Acknowledgment}
This work was supported by Guangzhou Municipal Science
and Technology Bureau under grant number 2025A03J3955. We would also like to extend our heartfelt thanks to Dr. Daniel Mortlock for his invaluable guidance and insightful discussions.

% \bibliographystyle{IEEEbib}
% \bibliography{icme2025references}

\end{document}